\documentclass[10pt,letterpaper,twocolumn]{article} 

\usepackage{ol}
\usepackage{hyperref}
\usepackage{amsmath}
\usepackage{bm}
\usepackage{amssymb}

\begin{document}

\twocolumn[ 

\title{Role of beam propagation in Goos-H\"{a}nchen and Imbert-Fedorov shifts
}


\author{A. Aiello,$^{1,*}$  J. P. Woerdman$^{1}$}
\address{$^1$Huygens Laboratory, Leiden University\\
P.O.\ Box 9504, 2300 RA Leiden, The Netherlands}
\address{$^*$Corresponding author: aiello@molphys.leidenuniv.nl}

\begin{abstract}
We derive the polarization-dependent displacements parallel and perpendicular to the plane of incidence, for a Gaussian light beam reflected from a planar interface, taking into account the propagation of the beam.  Using a classical-optics formalism we show that  beam propagation may greatly affect both Goos-H\"{a}nchen and Imbert-Fedorov shifts when the incident beam is  focussed.
%
\end{abstract}



 ] 

\maketitle 
It is known that the behavior of  bounded beams of light reflected from and transmitted through a planar interface differs from that exhibited by  plane waves, the latter being ruled by Snell's law and the Fresnel equations \cite{BandWBook}. For bounded beams diffractive corrections occur, the most prominent of which  the so-called
Goos-H\"{a}nchen  (GH) \cite{GH} and Imbert-Fedorov (IF) shifts \cite{CandI} of the beam, occurring in the directions parallel and perpendicular to the plane of incidence, respectively. In principle, both reflected and transmitted beams are subject to such shifts.
A great deal of literature exists about experimental \cite{Imbert,PillonEtAl} and theoretical
\cite{Nasalski2,BliokhPRL,LaiEtAl,Aiello} demonstrations of both GH and IF shifts but generally the effects of beam propagation  are not accounted for \cite{NoteCita}. A notable exception is a recent paper by Hosten and Kwiat  \cite{HostenandKwiat} where the authors report, among other issues, on a dramatic signal enhancement technique $(\sim 100 \times)$ for a \emph{quantum} version of the IF shift, the spin Hall effect of light (SHEL), based on beam propagation  \cite{NoteKwiatPol}.   The theoretical discussion in \cite{HostenandKwiat} uses the quantum formalism of weak measurements \cite{Vaid}, although the authors note that the \emph{beam propagation enhancement} (BPE) is essentially a classical phenomenon.

The purpose of this Letter is to present a purely classical analysis of the BPE; we feel that this is useful since a classical description will make this important technique, which allows sub-nm sensitivity \cite{HostenandKwiat},  better accessible to the metrology community.
Furthermore, our classical framework covers both the GH and the IF case, whereas the treatment in Ref.  \cite{HostenandKwiat} is restricted to the IF case only. A last, minor, difference between the present work and that of Hosten and Kwiat \cite{HostenandKwiat} is that the latter authors measure the beam that is transmitted across an air-glass
  interface, while we study the beam that is reflected by  such interface.
Since the transmission and reflection cases have a very similar mathematical
structure, all our main conclusions regarding reflective GH and IF shifts remain qualitatively valid for the transmission case.

We begin by considering optical reflection from a planar interface; Fig. 1 illustrates the coordinate system.
The $z$ axis of the laboratory Cartesian frame $(xyz)$ is normal to the planar interface $(z=0)$, that separates empty space (in practice air), where $z<0$,  from an optically dense region (either a dielectric or a metal \cite{Merano07}), where $z>0$.  We use a Cartesian frame $(x_i, y_i, z_i)$ attached to the incident beam and another one $(x_r, y_r, z_r)$ attached to the reflected beam.  Note that the coordinate $x_r$ is associated with the longitudinal GH shift, while $y_r$ is associated with the transverse IF shift.
%
%
%
Consider a monochromatic Gaussian beam of light  propagating in air parallel to the positive $z_i$ axis. The electric field amplitude of such beam can be written  as \cite{Aiello}
\begin{align}\nonumber
\mathbf{E}^\text{inc}
 \propto & \exp \left[i Z_i - \frac{ X_i^2 + Y_i^2 }{2\left(\Lambda + i Z_i \right)}\right] \\ \label{eq70}
 &\! \! \times \left( \hat{\mathbf{x}}_i f_P + \hat{\mathbf{y}}_i f_S - i \hat{\mathbf{z}}_i \frac{ f_P X_i +   f_S Y_i  }{\Lambda + i Z_i} \right),
\end{align}
where we have introduced the dimensionless variables $X_i = k x_i$,  $Y_i = k y_i$,  $Z_i = k z_i$, $\Lambda = k L$, and where the paraxial approximation corrected up to first order derivatives has been used \cite{HausandPan}.The  Gaussian amplitude of the  beam is characterized by the minimum waist $w_0$  located at $z_i =0$, and the Rayleigh range $L = k w_0^2/2$.
%
%
%
  \begin{flushleft}
  \begin{figure}[!hbr]
  \centerline{\includegraphics[width=6.5truecm]{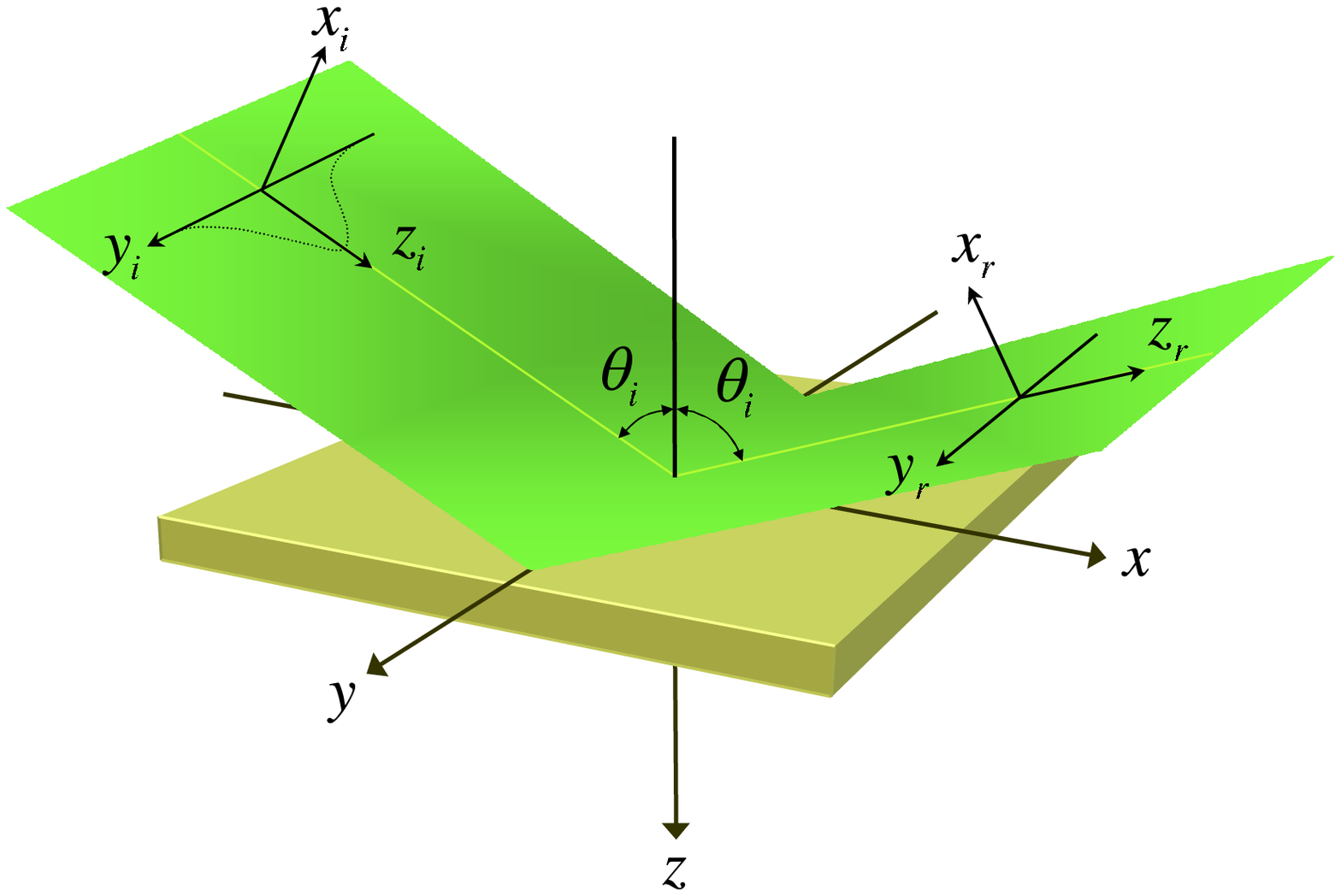}}
  \caption{(Color online) Geometry of beam reflection at the air-medium interface.}
  \end{figure}
  \end{flushleft}
%
%
The polarization of the  beam is determined by the complex-valued unit vector $\hat{\mathbf{f}} = \left( f_P  \hat{\mathbf{x}}_i + f_S  \hat{\mathbf{y}}_i\right)/\left( |f_P|^2 + |f_S|^2 \right)^{1/2}$, which corresponds experimentally to  a polarizer  perpendicular to the beam central wave vector $\mathbf{k}_i = k \hat{\mathbf{z}}_i$.
Across the interface $z=0$, the tangential components of the electric and magnetic field must be continuous. From this boundary condition and Eq. (\ref{eq70}),
the electric and magnetic fields of the reflected beam can be determined, under the same  conditions. A straightforward calculation \cite{Aiello} yields:
\begin{align}\nonumber
\mathbf{E}^\mathrm{ref}
 \propto   & \exp \left[i Z_r -\frac{X_r^2 + Y_r^2}{2 ( \Lambda + i Z_r )}\right]   \\ \nonumber
& \times \biggl\{ \Bigr. \hat{\mathbf{x}}_r \biggl[
  f_P  r_P \biggl(1 - i \frac{X_r}{\Lambda + i Z_r} \displaystyle{\frac{\partial \ln r_P}{\partial \theta_i} } \biggr) \\ \nonumber
& \qquad \qquad \! \! \! + i f_S \frac{Y_r}{\Lambda + i Z_r} \bigl( r_P  +  r_S  \bigl) \cot \theta_i
\biggr]
 \\ \nonumber
 &   + \hat{\mathbf{y}}_r  \biggl[
  f_S r_S \biggl(1
 - i \frac{X_r}{\Lambda + i Z_r} \displaystyle{\frac{\partial \ln  r_S}{\partial \theta_i} } \biggr)   \\ \nonumber
& \qquad \qquad \! \! \!  - i f_P \frac{Y_r}{\Lambda + i Z_r} \bigl( r_P  +  r_S  \bigl) \cot \theta_i
\biggr]
 \\ \label{eq80}
 &    - i \hat{\mathbf{z}}_r   \left(  \frac{r_P f_P X_r +  r_S f_S Y_r}{\Lambda + i Z_r} \right)
 \biggl. \biggr\},
\end{align}
where $ X_r  =   k x_r$, $ Y_r  =   k y_r$, $ Z_r  =   k z_r$, and $ r_A, \partial  r_A/ \partial \theta_i$, are the Fresnel reflection coefficients and their first derivatives evaluated at the ``central'' angle of incidence $\theta_i = \arccos( \mathbf{k}_i \cdot \hat{\mathbf{z}}/k )$, respectively. The index $A \in \{P,S \}$ is a label for a linearly polarized plane wave whose electric field vector is either parallel $(P)$ or perpendicular $(S)$ to the plane of incidence $(x,z)$, which is defined as the common plane of  the central wave vector $\mathbf{k}_i = k \hat{\mathbf{z}}_i$, and the normal to the interface $\hat{\mathbf{z}}$. In a similar manner the magnetic field  $\mathbf{B}^\mathrm{ref}$ of the reflected beam can be obtained and used to calculate the beam intensity  spatial profile $I (X_r,Y_r,Z_r)$ as the flux of the time averaged Poynting vector $\overline{\mathbf{S}} \propto {\mathrm{Re}\bigl(\mathbf{E}^\mathrm{ref} \times {\mathbf{B}^\mathrm{ref}}^* \bigr)}$ through a surface perpendicular to the central direction of propagation $\hat{\mathbf{z}}_r$: $I (X_r,Y_r,Z_r) \propto \overline{\mathbf{S}}  \cdot \hat{\mathbf{z}}_r$.
At any given plane $Z_r = \text{const.}$, the intensity $I(X_r,Y_r,Z_r)$  can be considered as the distribution of the ``quasi-Gaussian'' variables $\mathbf{X} = X_r \hat{\mathbf{x}}_r + Y_r \hat{\mathbf{y}}_r$ with means $\mathbf{M} = \left\langle X_r \right\rangle \hat{\mathbf{x}}_r + \left\langle Y_r \right\rangle \hat{\mathbf{y}}_r$, where %
 %
\begin{align}\label{eq100}
\mathbf{M} =  \frac{\int \! \! \int  \mathbf{X} \, I(X_r,Y_r,Z_r) \text{d}X_r \text{d}Y_r}{ \int \! \!  \int I(X_r,Y_r,Z_r) \text{d}X_r \text{d}Y_r}
\end{align}
determines the centroid of the reflected beam \cite{Li}. As the result of a straightforward calculation, one obtains for the GH shift
\begin{align}\nonumber
\left\langle X_r \right\rangle
 = &   \frac{ \varphi_P R_P^2 a_P^2 + \varphi_S R_S^2  a_S^2 }{R_P^2  a_P^2 + R_S^2  a_S^2}\\ \label{GH}
 & - \frac{Z_r}{\Lambda} \frac{ \rho_P R_P^2 a_P^2 + \rho_S R_S^2  a_S^2 }{R_P^2  a_P^2 + R_S^2  a_S^2},
\end{align}
and for the IF shift
\begin{align}\nonumber
\left\langle Y_r \right\rangle =  & - \frac{  a_P a_S \cot \theta_i   }{R_P^2  a_P^2 + R_S^2  a_S^2} \biggl\{ \Bigl[ \bigl( R_P^2 + R_S^2  \bigr)  \sin \eta   \\ \nonumber
&   + 2 R_P R_S  \sin (\eta - \phi_P + \phi_S ) \Bigr]\\ \label{IF}
& - \frac{Z_r}{\Lambda}\bigl( R_P^2 - R_S^2  \bigr)  \cos \eta \biggr\},
\end{align}
where $r_A \equiv R_A \exp(i \phi_A), \; A \in \{ P, S\}$, $f_P = a_P \in \mathbb{R}$, $f_S = a_S \exp(i \eta)$, and $\rho_A = \text{Re} \left( {\partial \ln r_A}/{\partial \theta_i} \right)$, $\varphi_A = \text{Im} \left( {\partial \ln r_A}/{\partial \theta_i} \right)$.

Equations (\ref{GH},\ref{IF}) are the first main result of this Letter. They give both the GH and the IF shift as functions of the beam propagation distance $Z_r$.  For a well collimated beam, the condition $Z_r/\Lambda \ll 1$ is trivially satisfied at optical frequencies and  the $Z_r$-independent terms in both Eqs.(\ref{GH}-\ref{IF}) are dominant.
Such terms represent the ``traditional''  GH and IF shifts. In fact, when $Z_r=0$, Eq. (\ref{GH}) is a straightforward generalization of the well known Artmann formula \cite{Artmann}; and  Eq. (\ref{IF}) is in agreement with Bliokh and Bliokh \cite{BliokhPRE}.
However, for a focussed beam the condition $Z_r/\Lambda \gg 1$  may hold, and the $Z_r$-dependent terms become relevant.
For example, if a typical He-Ne laser operating at wavelength $\lambda$ of $633$ nm with a minimum waist $w_1$ of $1.5$ mm, is focussed by a lens with focal length $f$ of $100$ mm, a waist $w_0 = \lambda f / (\pi w_1) \simeq 13$ $\mu$m is produced. If this beam propagates over a distance of $250$ mm from the lens, we easily obtain $Z_r/\Lambda \simeq 168$. Thus, depending on the actual experimental conditions, either $Z_r$-dependent or $Z_r$-independent terms in Eqs. (\ref{GH},\ref{IF}) may be dominant, and, as a consequence, the measured GH and IF shifts will dramatically change.

In order to show the connections between the results above and the ones presented in Ref. \cite{HostenandKwiat}, we must take a step backward and calculate, as an illustrative  example, the quantity $\hat{\mathbf{x}}_r \cdot \mathbf{E}^\mathrm{ref}$ evaluated for $a_S=0$ on the plane of incidence $Y_r =0$. This corresponds to an experimental configuration apt to measure the GH shift of a $P$-polarized beam. From Eq. (\ref{eq80}) it readily follows
\begin{align} \nonumber
\hat{\mathbf{x}}_r \! \cdot \! \mathbf{E}^\mathrm{ref} \! \propto  & \exp\biggl[ -\frac{X_r^2}{2 ( \Lambda + i Z_r ) } \biggr] \! \! \left(
1 - i \frac{X_r}{\Lambda + i Z_r} \displaystyle{\frac{\partial \ln r_P}{\partial \theta_i} } \right) \\ \label{eq110}
\sim & \exp \biggl[{ -\frac{\left( X_r - \varphi_P + i \rho_P \right)^2}{2 ( \Lambda + i Z_r ) } }   \biggr].
\end{align}
A careful inspection of the equation above, shows that it represents a Gaussian beam tilted clockwise  by an angle $\rho_P/\Lambda$ with respect to the axis $\hat{\mathbf{z}}_r$, and displaced by $\varphi_P/k$ along  the axis $\hat{\mathbf{x}}_r$. In other words, for a Gaussian beam an \emph{imaginary} position shift is equivalent to an angular tilt, while a \emph{real} one represents a spatial shift. It is easy to see that the complex-valued nature of the shift, controls its behavior under beam propagation. To demonstrate this, we note that calculation of the centroid of the distribution $|\hat{\mathbf{x}}_r \cdot \mathbf{E}^\mathrm{ref}|^2$ yields
 \begin{align} \label{eq115}
\langle X_r \rangle = \varphi_P - ({Z_r }/{ \Lambda})\rho_P,
\end{align}
%
%
which coincides with Eq. (\ref{GH}) evaluated for $a_S =0$. This expression of  $\langle X_r \rangle$ shows that only the imaginary part $\rho_P$ of the  complex shift $\varphi_P - i \rho_P$  couples to the coordinate $Z_r$ and it is enhanced by a factor $Z_r/\Lambda$ as the reflected beam  propagates along the optical distance $Z_r$.
As it will be shown below, such factor $Z_r/\Lambda$  coincides with the propagation enhancement factor $F$ of Ref. \cite{HostenandKwiat} which lies at the core of their signal enhancement technique.
Although in the reasoning above we have considered a specific example, it is not difficult to realize that the conclusions reached  are perfectly general; in particular, positioning of the waist of the Gaussian beam at the interface is not essential (see also Ref. \cite{HostenandKwiat}). This is the second main result of our Letter.

We conclude this Letter by demonstrating that, as anticipated, the term $Z_r/\Lambda$ in our Eq. (\ref{eq115}) is coincides with the propagation enhancement factor $F$ of Hosten and Kwiat \cite{HostenandKwiat}.
 According to their  scheme, the incident beam is first pre-selected in the $P$ polarization state, namely $f_P=1,\,f_S=0$, and then post-selected (after reflection by an air-glass interface) in the $\hat{\mathbf{v}}$ polarization state, where  $\hat{\mathbf{v}} = \hat{\mathbf{x}}_r \sin \Delta  + \hat{\mathbf{y}}_r \cos \Delta$, and $\Delta = \pm |\Delta|$. In addiction, after reflection the beam is observed along the transverse plane $X_r =0$. Thus, the relevant amplitude of the reflected field can be written as
\begin{align}\nonumber
 \hat{\mathbf{v}}
\! \cdot \! \mathbf{E}^\mathrm{ref} \Bigl|_{X_r =0} \propto & \exp \left[{-\frac{Y_r^2}{2 ( \Lambda + i Z_r )}}\right]\left(
1 + i \frac{Y_r  D \cot \Delta}{\Lambda + i Z_r} \right)\\ \label{eq120}
\sim & \exp \biggl[{ -\frac{\left( Y_r - i D \cot \Delta \right)^2}{2 ( \Lambda + i Z_r ) } }   \biggr].
\end{align}
where  the purely imaginary term $i D \cot \Delta$ is described in Ref.  \cite{HostenandKwiat} as the product  of the SHEL-induced photon displacement $D =
\left( 1  +  {r_S}/{r_P}  \right)\cot \theta_i$, and the weak value of the photon spin component $i \cot \Delta$.
However, by comparing Eq. (\ref{eq120}) with Eq. (\ref{eq110}), it is clear that at a classical level such imaginary shift amounts to a tilt by the small angle $D \cot \Delta/\Lambda$, of the reflected beam. From the discussion above we know that an imaginary shift  couples with coordinate $Z_r$ and, therefore,  increases as the beam propagates. In fact,  calculation of the centroid of the distribution $\left| \hat{\mathbf{v}}
\! \cdot \! \mathbf{E}^\mathrm{ref} \right|_{X_r =0}|^2$ yields
\begin{align}\label{eq130}
 \langle Y_r \rangle  =  ({Z_r}/{\Lambda}) {{D}} \cot \Delta, \quad  \langle Y_r^2 \rangle  =  ({\Lambda^2 + Z_r^2})/({2 \Lambda}) ,
\end{align}
where $Z_r/ \Lambda \equiv F$ ($ \simeq 156$ in Ref. \cite{HostenandKwiat}) is just the  propagation enhancement factor. This last equality can be easily proved by combining the two formulas in Eq. (\ref{eq130}) to obtain, in terms of dimensional variables,
\begin{align}\label{eq140}
 \langle y_r \rangle  =  \frac{2 k\langle y_r^2 \rangle }{R(z)} \delta \cot \Delta \simeq \frac{4 \pi \langle y_r^2 \rangle }{z \lambda} \delta \cot \Delta,
\end{align}
where $\delta = D/k$, $R(z) = (z^2 + L^2)/z \simeq z$ is the radius of curvature of the beam wavefront at the lens L2 position \cite{NoteKwiat2}, and the last approximate equality holds in the experimental conditions of Ref.  \cite{HostenandKwiat}, where $z/L \gg 1$. From Eq. (\ref{eq140}) it readily follows that $F = {4 \pi \langle y_r^2 \rangle }/{(z \lambda)}$, in agreement with Eq. (4) in Ref. \cite{HostenandKwiat}.

In  summary, we have furnished analytic expressions, based upon classical optics, for both the GH and the IF shifts, as functions of the beam propagation distance. These  give in a natural way the dramatic changes of the  GH and  IF displacements induced by using a focussed beam. Moreover, from the analysis of such expressions, we derived a fully classical interpretation of a very recently introduced
signal enhancement technique,  employed to measure the spin Hall effect of light \cite{HostenandKwiat}.
%

We acknowledge Michele Merano  for  useful discussions. This
project is supported by FOM.

%

\begin{thebibliography}{10}
\bibitem{NoteCita}
The literature on these topics is so vast that it is impossible to cite the
  complete bibliography. As an example, see
  \cite{Imbert,PillonEtAl,Nasalski2,BliokhPRL,LaiEtAl} and references therein.
\bibitem{NoteKwiatPol}
It should be noted, however, that in Ref. \cite{HostenandKwiat} an additional
  signal enhancement technique based upon polarization control, is employed. In
  this Letter we do not cover such issue since the beam propagation factor is
  independent of polarization.
\bibitem{NoteKwiat2}
Note that in Ref. \cite{HostenandKwiat} the authors use a lens L2 with focal
  length $z_\text{eff}$ to collimate the beam so that $\langle y_r \rangle$
  remains constant after passing L2.
\end{thebibliography}
%

\end{document}